\documentclass[showpacs,amsmath,amssymb,nofootinbib,superscriptaddress]{revtex4}

\usepackage{graphicx}
\usepackage{epsfig}
\usepackage{bm}
\usepackage{amsfonts}
\usepackage{amssymb}
\usepackage{color}
\usepackage{float}
\usepackage{amsmath}                  
\usepackage{dcolumn}
\usepackage{hyperref}
\usepackage{array}
\usepackage{ulem}
\usepackage{xcolor}
\usepackage{lscape}
\usepackage{tensor}
\renewcommand{\(}{\left(}
\renewcommand{\)}{\right)}

\usepackage{graphicx,subfigure,bm,color,psfrag,hyperref}

\usepackage{mathtools}

\usepackage{amsmath,amsthm}
\usepackage{amsfonts}
\usepackage{amssymb}
\usepackage{verbatim}

\newcommand{\lin}{\\[7pt]}

\newcommand{\der}[2]{\dfrac{d#1}{d#2}}
\newcommand{\pder}[2]{\dfrac{\partial#1}{\partial#2}}
\newcommand{\pdder}[3]{\dfrac{\partial^2 #1}{\partial #2 \partial #3}}

\newcommand{\dder}[2]{\dfrac{\delta#1}{\delta#2}}
\newcommand{\pdot}[1]{\dot{\partial}_{#1}}

\newcommand{\D}{\mathcal{D}}
\newcommand{\Gd}{\mathcal{G}}

\newcommand{\R}{\mathcal{R}}
\newcommand{\de}{\mathrm{d}}

\begin{document}
	
	\title{Gravitational Field on the Lorentz Tangent Bundle: Generalized Paths and Field Equations}
	
	\author{A. Triantafyllopoulos}
	\email{alktrian@phys.uoa.gr}
	\affiliation{Section of Astrophysics, Astronomy and Mechanics, Department of Physics,  
		National and Kapodistrian University of Athens, Panepistimiopolis 15784,
		Athens, Greece} 
	
	\author{E. Kapsabelis}
	\email{manoliskapsabelis@yahoo.gr}
	\affiliation{Section of Astrophysics, Astronomy and Mechanics, Department of Physics,  
		National and Kapodistrian University of Athens, Panepistimiopolis 15784,
		Athens, Greece} 
	
	\author{P. C. Stavrinos}
	\email{pstavrin@math.uoa.gr}
	\affiliation{Department of Mathematics, National and Kapodistrian University of Athens,
		Panepistimiopolis 15784, Athens, Greece}

	\begin{abstract}
		We investigate the dynamics of gravitational field and particles in a generalized framework of a Lorentz tangent bundle. By variating an appropriate action for each case, we obtain generalized forms of paths and generalized field equations for a Sasaki type metric. We show that Stokes theorem is modified with respect to general relativity due to local anisotropy and the presence of a nonlinear connection which induces an adapted basis in our space.
	\end{abstract}
	
	
	\maketitle
\section{Introduction}
The gravitational field is an essential property of space-time and it is intimately related to the mass in the universe. The fundamental ingredients of the gravitational field are not yet completely known in physics.

In the last two decades there have been attempts by Theoretical Physicists and Cosmologists to discover observational imprints of the gravitational field which is spread out in more than four dimensions \cite{perez2005introduction,Chakraborty:2017qve,SatheeshKumar:2006ac,emparan2008black,freedman1985}. These considerations lead to the increase of degrees of freedom of the gravitational field. Alternatively, gravity can also be studied in a different way in the framework of an 8-dimensional Lorentz tangent bundle or a vector bundle which includes the observer (velocity/tangent vector) with extra internal/dynamical degrees of freedom \cite{stavrinos-ikeda 1999,Triantafyllopoulos:2018bli,Hohmann:2018rpp,Pfeifer:2015tua,Vacaru:2011pb,Vacaru:2009ye}.  In some cases the gravitational field has been considered in higher dimensions in relation to the Kaluza-Klein theory, general relativity modifications, string theory or the braneworld models \cite{Overduin:1998pn,Green:2012oqa,Maartens:2010ar}. This approach is provided by a generalized geometrical structure in the dynamics of the gravitational field and the field equations.

Furthermore, locally anisotropic structures of space-time in higher dimensions have been studied in the framework of Finsler and Finsler-like geometries \cite{Minas:2019urp,Ikeda:2019ckp,Stavrinos:2012kv}. The development of research for the evolution of the universe can be combined with a locally anisotropic structure of the Finslerian gravitational field. Finsler-gravity models allow intrinsically local anisotropies including vector variables $y^\mu=\frac{dx^\mu}{d\tau},\, (\mu=0,1,2,3)$, in the framework of a vector/tangent (Lorentz) bundle \cite{stavrinos-ikeda 1999,Triantafyllopoulos:2018bli,Hohmann:2018rpp,Pfeifer:2015tua,Stavrinos:2012ty,Stavrinos:2014apa}. The $y-$dependence essentially characterizes the Finslerian gravitational field and has been combined with the concept of anisotropy and the broken Lorentz symmetry which causes the deviation from Riemannian geometry, since the latter cannot explain completely all the gravitational effects in the universe.  In this approach
remarkable efforts have been made based on the Finslerian structure of
space-time. There have been proposed frameworks for testing definite Finsler modifications of general relativity which are in agreement with observational effects. Corrections have been proposed for the paths of perihelion of
Mercury, gravitational lensing, redshift e.t.c. \cite{Pfeifer:2015tua,aringazin1985,Laemmerzahl:2015rya,Hohmann:2016pyt}. Therefore, the consideration of Finsler geometry as a candidate for studying gravitational theories ensures that matter dynamics take place \cite{papagiannopoulos-basilakos-paliathanasis-savvidou-stavrinos 2017,kostelecky 2004}. Finsler gravity and cosmology models were developed by extending geometrical and physical ideas and have been related to quantum gravity and modified gravity theories, e.g. \cite{perelman2020borns,Fuster:2015tua,Vacaru:2010fa,Kostelecky:2010hs}.

A basic characteristic concept in such theories is the nonlinear connection which connects external and internal structures of space-time in higher order dimensions. The derived field equations of these gravitational and cosmological models constitute  a generalized form of field equations with more than one curvature and energy momentum tensors. It is fundamental for studying locally-anisotropic space on the tangent bundle of a 4-dimensional space-time manifold with internal structure \cite{Miron:1994nvt,Vacaru:2005ht}.

In the present work we study the dynamics of the gravitational field. In section \ref{sec: paths} we study the generalized equations of paths in the framework of a Lorentz tangent bundle. In section \ref{sec: field equations} we derive the field equations in a sufficiently generalized form by using torsion terms. In section \ref{sec: conclusion} some concluding remarks are given.

\section{Preliminaries}\label{sec: preliminaries}
The natural background space for a locally anisotropic gravity is the tangent bundle of a differentiable Lorentzian space-time manifold called a Lorentz Tangent Bundle (we will refer to it as $ TM $ hereafter) \cite{Miron:1994nvt,Vacaru:2005ht}. $TM$ is itself an 8-dimensional differentiable manifold, so we can define coordinate charts and tensors on it in the usual way. $TM$ is equipped with local coordinates $\{\mathcal{U}^A \} = \{x^\mu,y^\alpha\}$ where $x^\mu$ are the local coordinates on the base manifold $M$ around $\pi(\sigma)$, $\sigma \in TM$, and $y^\alpha$ are the coordinates on the fiber. The range of values for the indices is $\kappa,\lambda,\mu,\nu,\ldots = 0,\ldots,3$ and $\alpha,\beta,\ldots,\theta = 4,\ldots,7$.

The adapted basis on the total space $TTM$ is defined as $\{E_A\} = \,\{\delta_\mu,\dot\partial_\alpha\} $ where
\begin{equation}
\delta_\mu = \dfrac{\delta}{\delta x^\mu}= \pder{}{x^\mu} - N^\alpha_\mu(x,y)\pder{}{y^\alpha} \label{delta x}
\end{equation}
and
\begin{equation}
\dot \partial_\alpha = \pder{}{y^\alpha}
\end{equation}
where $N^\alpha_\mu$ are the components of a nonlinear connection. The curvature of the nonlinear connection is defined as
\begin{equation}\label{Omega}
\Omega^\alpha_{\nu\kappa} = \dder{N^\alpha_\nu}{x^\kappa} - \dder{N^\alpha_\kappa}{x^\nu}
\end{equation}
The nonlinear connection induces a split of the total space $TTM$ into a horizontal distribution $T_HTM$ and a vertical distribution $T_VTM$. The above-mentioned split is expressed with the Whitney sum:
\begin{equation}
TTM = T_HTM \oplus T_VTM
\end{equation}
The horizontal distribution or h-space is spanned by $\delta_\mu$, while the vertical distribution or v-space is spanned by $\pdot \alpha$. Under a local coordinate transformation, the adapted basis vectors transform as:
\begin{equation}\label{h basis transformation}
\delta_{\mu'} = \pder{x^\mu}{x^{\mu'}}\delta_\mu \quad,\quad \pdot{\alpha'} = \pder{x^\alpha}{x^{\alpha'}}\pdot{\alpha}
\end{equation}
The adapted dual basis of the adjoint total space $T^*TM$ is $ \{E^A \}  = \{\de x^\mu, \delta y^\alpha\}$ with the definition
\begin{equation}
\delta y^\alpha = \mathrm{d}y^\alpha + N^\alpha_\nu\mathrm{d}x^\nu \label{delta y}
\end{equation}
The transformation rule for $ \{\de x^\mu, \delta y^\alpha\} $ is:
\begin{equation}
\de x^{\mu'} = \pder{x^{\mu'}}{x^\mu}\de x^\mu \quad,\quad \delta y^{\alpha'} = \pder{x^{\alpha'}}{x^\alpha}\delta y^\alpha
\end{equation}
The bundle $TM$ is equipped with a distinguished metric ($d-$metric) $\Gd$:
\begin{equation}
\mathcal{G} = g_{\mu\nu}(x,y)\,\mathrm{d}x^\mu \otimes \mathrm{d}x^\nu + v_{\alpha\beta}(x,y)\,\delta y^\alpha \otimes \delta y^\beta \label{bundle metric}
\end{equation}
where the h-metric $g_{\mu\nu}$ and v-metric $v_{\alpha\beta}$ are defined to be of Lorentzian signature $(-,+,+,+)$. A tangent bundle equipped with such a metric will be called a Lorentz tangent bundle. In some cases, the following homogeneity conditions will be assumed: $g_{\mu\nu}(x,ky) = g_{\mu\nu}(x,y), v_{\alpha\beta}(x,ky) = v_{\alpha\beta}(x,y), k>0$. When these conditions are met, the following relations hold:
\begin{align}
g_{\alpha\beta} = \mathrm{sgn}(g)\frac{1}{2}\pdder{F^2_g}{y^\alpha}{y^\beta} \\
v_{\alpha\beta} = \mathrm{sgn}(v)\frac{1}{2}\pdder{F^2_v}{y^\alpha}{y^\beta}
\end{align}
where $ g_{\alpha\beta} = \tilde{\delta_\alpha^\mu}\tilde{\delta_\beta^\nu}g_{\mu\nu} $\footnote{The generalized Kronecker symbols are defined as: $\tilde{\delta_\alpha^\mu} = \tilde\delta^\alpha_\mu = 1$ for $a=\mu+4$ and equal to zero otherwise.}, $\mathrm{sgn}(g)$ is the sign of $g_{\alpha\beta}(x,y)y^\alpha y^\beta$, $ \mathrm{sgn}(v) $ is the sign of $v_{\alpha\beta}(x,y)y^\alpha y^\beta$ and
\begin{align}
F_g(x,y) = \sqrt{|g_{\alpha\beta}(x,y)y^\alpha y^\beta|} \\
F_v(x,y) = \sqrt{|v_{\alpha\beta}(x,y)y^\alpha y^\beta|}
\end{align}
From the above relations, the following conditions become obvious:
\begin{enumerate}
	\item $F_m$, $m=g,v$, is continuous on $TM$ and smooth on  $ \widetilde{TM}\equiv TM\setminus \{0\} $ i.e. the tangent bundle minus the null set $ \{(x,y)\in TM | F_m(x,y)=0\}$ \label{finsler field of definition}
	\item $ F_m $ is positively homogeneous of first degree on its second argument:
	\begin{equation}
	F_m(x^\mu,ky^\alpha) = kF_m(x^\mu,y^\alpha), \qquad k>0 \label{finsler homogeneity}
	\end{equation}
	\item The form \begin{equation}f_{\alpha\beta}(x,y) = \dfrac{1}{2}\pdder{F_m^2}{y^\alpha}{y^\beta} \label{finsler metric} \end{equation} defines a non-degenerate matrix: \label{finsler nondegeneracy}
	\begin{equation}
	\det\left[f_{\alpha\beta}\right] \neq 0 \label{finsler nondegenerate}
	\end{equation}
\end{enumerate}

In this work, we consider a distinguished connection ($d-$connection) $ {D} $ on $TM$. This is a linear connection with coefficients $\{\Gamma^A_{BC}\} = \{L^\mu_{\nu\kappa}, L^\alpha_{\beta\kappa}, C^\mu_{\nu\gamma}, C^\alpha_{\beta\gamma} \} $ which preserves by parallelism the horizontal and vertical distributions:
	\begin{align}
	{D_{\delta_\kappa}\delta_\nu = L^\mu_{\nu\kappa}(x,y)\delta_\mu}\, \quad &,\quad D_{\pdot{\gamma}}\delta_\nu = C^\mu_{\nu\gamma}(x,y)\delta_\mu \label{D delta nu} \lin
	{D_{\delta_\kappa}\pdot{\beta} = L^\alpha_{\beta\kappa}(x,y)\pdot{\alpha}} \quad &, \quad D_{\pdot{\gamma}}\pdot{\beta} = C^\alpha_{\beta\gamma}(x,y)\pdot{\alpha} \label{D partial b}
	\end{align}
	From these, the definitions for partial covariant differentiation follow as usual, e.g. for $X \in TTM$ we have the definitions for covariant h-derivative
	\begin{equation}
	X^A_{|\nu} \equiv D_\nu\,X^A \equiv \delta_\nu X^A + L^A_{B\nu}X^B \label{vector h-covariant}
	\end{equation}
	and covariant v-derivative
	\begin{equation}
	X^A|_\beta \equiv D_\beta\,X^A \equiv \dot{\partial}_\beta X^A + C^A_{B\beta}X^B \label{vector v-covariant}
	\end{equation}
	
A $d-$connection can be uniquely defined given that the following conditions are satisfied:
\begin{itemize}
	\item The $d-$connection is metric compatible
	\item Coefficients $L^\mu_{\nu\kappa}, L^\alpha_{\beta\kappa}, C^\mu_{\nu\gamma}, C^\alpha_{\beta\gamma} $ depend solely on the quantities $g_{\mu\nu}$, $v_{\alpha\beta}$ and $N^\alpha_\mu$
	\item Coefficients $L^\mu_{\kappa\nu}$ and $ C^\alpha_{\beta\gamma} $ are symmetric on the lower indices, i.e.  $L^\mu_{[\kappa\nu]} = C^\alpha_{[\beta\gamma]} = 0$
\end{itemize}
	We use the symbol $\mathcal D$ instead of $D$ for a connection satisfying the above conditions, and call it a canonical and distinguished $d-$connection. Metric compatibility translates into the conditions:
	\begin{equation}
	\mathcal D_\kappa\, g_{\mu\nu} = 0, \quad \mathcal D_\kappa\, v_{\alpha\beta} = 0, \quad\mathcal D_\gamma\, g_{\mu\nu} = 0, \quad\mathcal D_\gamma\, v_{\alpha\beta} = 0
	\end{equation}
	The coefficients of canonical and distinguished $d-$connection are
	\begin{align}
	L^\mu_{\nu\kappa} & = \frac{1}{2}g^{\mu\rho}\left(\delta_kg_{\rho\nu} + \delta_\nu g_{\rho\kappa} - \delta_\rho g_{\nu\kappa}\right) \label{metric d-connection 1}  \\
	L^\alpha_{\beta\kappa} & = \dot{\partial}_\beta N^\alpha_\kappa + \frac{1}{2}v^{\alpha\gamma}\left(\delta_\kappa v_{\beta\gamma} - v_{\delta\gamma}\,\dot{\partial}_\beta N^\delta_\kappa - v_{\beta\delta}\,\dot{\partial}_\gamma N^\delta_\kappa\right) \label{metric d-connection 2}  \\
	C^\mu_{\nu\gamma} & = \frac{1}{2}g^{\mu\rho}\dot{\partial}_\gamma g_{\rho\nu} \label{metric d-connection 3} \\
	C^\alpha_{\beta\gamma} & = \frac{1}{2}v^{\alpha\delta}\left(\dot{\partial}_\gamma v_{\delta\beta} + \dot{\partial}_\beta v_{\delta\gamma} - \dot{\partial}_\delta v_{\beta\gamma}\right) \label{metric d-connection 4}
	\end{align}
	
	Curvature and torsion in $TM$ can be defined as multilinear maps:
	\begin{equation}
	\mathcal{R}(X,Y)Z = [\mathcal{D}_X,\mathcal{D}_Y]Z - \mathcal{D}_{[X,Y]}Z \label{Riemann tensor TM}
	\end{equation}
	and
	\begin{equation}
	\mathcal{T}(X,Y) = \mathcal{D}_XY - \mathcal{D}_YX - [X,Y] \label{torsion TM}
	\end{equation}
	where $X,Y,Z \in TTM$.
	We use the definitions
	\begin{align}
	\mathcal{R}(\delta_\lambda,\delta_\kappa)\delta_\nu = R^\mu_{\nu\kappa\lambda}\delta_\mu \label{R curvature components} \lin
	\mathcal{R}(\pdot{\delta},\pdot{\gamma})\pdot{\beta} = S^\alpha_{\beta\gamma\delta}\dot{\partial}_\alpha \label{S curvature components}
	\end{align}
	
	\begin{align}
	\mathcal{T}(\delta_\kappa,\delta_\nu) = & \mathcal{T}^\mu_{\nu\kappa}\delta_\mu + \mathcal{T}^\alpha_{\nu\kappa}\pdot{\alpha} \label{torsion components 1} \lin
	\mathcal{T}(\pdot{\gamma},\pdot{\beta}) = & \mathcal{T}^\mu_{\beta\gamma}\delta_\mu + \mathcal{T}^\alpha_{\beta\gamma}\pdot{\alpha} \label{torsion components 3}
	\end{align}
	The h-curvature tensor of the $d-$connection in the adapted basis and the corresponding h-Ricci tensor have, respectively, the components
	\begin{gather}
	R^\mu_{\nu\kappa\lambda} = \delta_\lambda L^\mu_{\nu\kappa} - \delta_\kappa L^\mu_{\nu\lambda} + L^\rho_{\nu\kappa}L^\mu_{\rho\lambda} - L^\rho_{\nu\lambda}L^\mu_{\rho\kappa} + C^\mu_{\nu\alpha}\Omega^\alpha_{\kappa\lambda} \label{R coefficients 1}\lin
	R_{\mu\nu} = R^\kappa_{\mu\nu\kappa} = \delta_\kappa L^\kappa_{\mu\nu} - \delta_\nu L^\kappa_{\mu\kappa} + L^\rho_{\mu\nu}L^\kappa_{\rho\kappa} \nonumber
	- L^\rho_{\mu\kappa}L^\kappa_{\rho\nu} + C^\kappa_{\mu\alpha}\Omega^\alpha_{\nu\kappa} \label{d-ricci 1}
	\end{gather}
	The v-curvature tensor of the $d-$connection in the adapted basis and the corresponding v-Ricci tensor have, respectively, the components
	\begin{align}
	S^\alpha_{\beta\gamma\delta} & = \pdot{\delta} C^\alpha_{\beta\gamma} - \pdot{\gamma}C^\alpha_{\beta\delta} + C^\epsilon_{\beta\gamma}C^\alpha_{\epsilon\delta} - C^\epsilon_{\beta\delta}C^\alpha_{\epsilon\gamma} \label{S coefficients 2} \lin
	S_{\alpha\beta} & = S^\gamma_{\alpha\beta\gamma} = \pdot{\gamma}C^\gamma_{\alpha\beta} - \pdot{\beta}C^\gamma_{\alpha\gamma} + C^\epsilon_{\alpha\beta}C^\gamma_{\epsilon\gamma} - C^\epsilon_{\alpha\gamma}C^\gamma_{\epsilon\beta} \label{d-ricci 4}
	\end{align}
	The generalized Ricci scalar curvature in the adapted basis is defined as
	\begin{equation}
	\R = g^{\mu\nu}R_{\mu\nu} + v^{\alpha\beta}S_{\alpha\beta} = R+S \label{bundle ricci curvature}
	\end{equation}
	where
	\begin{align}
	R=g^{\mu\nu}R_{\mu\nu} \quad,\quad
	S=v^{\alpha\beta}S_{\alpha\beta} \label{hv ricci scalar}
	\end{align}
	
	\section{Generalized equations of paths from a variational principle}\label{sec: paths}
	
	In this section, we investigate the curves of point particles on the Lorentz tangent bundle with a Lagrangian function of the form $L(x,\dot x,y)$. Our goal is to examine under which conditions there is a relation between $y$ and $\dot x$ directional variables.
	
	In many cases, if someone wants to describe curves $(x(s),y(s))$ of point particles with mass $m$ in Finsler or Finsler-like geometry, he uses the Lagrangian $L(x,\dot x) = -m\left(g_{\mu\nu}(x,y)\dot x^\mu \dot x^\nu\right)^{1/2}$ with $\dot x^\mu = d x^\mu/ds$. In this case, variation of the action $K = \int L ds$ with respect to $x^\mu$ gives the geodesics equation
	\begin{equation}\label{geodesics}
	\ddot x^\mu + \gamma^\mu_{\kappa\lambda}(x,y)\dot x^\kappa \dot x^\lambda = 0
	\end{equation}
	where $\gamma^\mu_{\kappa\lambda}(x,y)$ are the Christoffel symbols for the metric $g_{\mu\nu}(x,y)$ and we assumed the homogeneity condition $g_{\mu\nu}(x,ky) = g_{\mu\nu}(x,y)$ for $k>0$. Equation \eqref{geodesics} does not fully define a curve on $TM$, since $y(s)$ remains undefined.  Usually, in gravitational models on a Lorentz tangent bundle, the fiber coordinates $y^\alpha$ represent 4-velocity components $ \dot x^\nu $.
	
	A condition that relates $y$ and $\dot x$ is of the form
	\begin{equation}\label{y=xdotin}
	y^\alpha = \tilde \delta^\alpha_\nu \dot x^\nu
	\end{equation}
	and so the curves on $TM$ are fully described by relations \eqref{geodesics} and \eqref{y=xdotin}\footnote{The generalized Kronecker symbols are defined as: $\tilde\delta^\alpha_\mu = 1$ for $a=\mu+4$ and zero otherwise, $ \mu = 0,1,2,3$.}. We note that relation \eqref{y=xdotin} is not derived from a variational principle.
	
	On the other hand, there have been some studies to derive geodesics and their deviation equations for both $x$ and $y$ from an action on the tangent bundle \cite{asanov1991finslerian,balan1999weak}, however, in the resulting equations the fiber elements do not have any apparent relation to 4-velocities.
	
	\subsection{Curves from a total Lagrangian}
	
	In the following, we present a different approach for the derivation of curves by a variational principle. We choose a Lagrangian function $L$ for the fields $\dot x^\mu$ and $y^\alpha$ with the following properties:
	\begin{itemize}
		\item $ L $ includes a free term for the field $\dot x^\mu$. A suitable choice is the norm of $\dot x^\mu$ scaled by some constant $a$, namely $ a g_{\mu\nu} \dot x^\mu \dot x^\nu $, with $g_{\mu\nu}(x,y) $ the h-metric of \eqref{bundle metric}.
		\item $ L $ includes a free term for the field $ y^\alpha $. A natural choice would be the norm of $y$ scaled by some constant $c$, namely $cv_{\alpha\beta}y^\alpha y^\beta$, with $v_{\alpha\beta}(x,y)$ the v-metric of \eqref{bundle metric}.
		\item $ L $ includes an interaction term between $ \dot x^\mu$  and $ y^\alpha $. We assume that the interaction is dependent on the internal space v-metric $v_{\alpha\beta}$ so we choose $b\tilde \delta^\alpha_\mu v_{\alpha\beta}\dot x^\mu y^\beta$, with b constant.
	\end{itemize}
	We write the full Lagrangian as:
	\begin{equation}\label{full lagrangian}
	L(x,\dot x,y) = \left(a g_{\mu\nu}\dot x^\mu \dot x^\nu + b \tilde \delta^\alpha_\mu v_{\alpha\beta}\dot x^\mu y^\beta + c v_{\alpha\beta}y^\alpha y^\beta \right)^{1/2}
	\end{equation}
	The Euler-Lagrange equations for \eqref{full lagrangian} are
	\begin{align}
	\pder{L}{x^\kappa} - \der{}{s}\left(\pder{L}{\dot x^\kappa}\right) = 0 \label{E-L x} \\
	\pder{L}{y^\gamma}  - \der{}{s}\left(\pder{L}{\dot y^\gamma}\right) = 0 \label{E-L y}
	\end{align}
	Now, relation \eqref{E-L y} gives
	\begin{align}
	& v_{\alpha\gamma}\left( b\tilde\delta^\alpha_\mu \dot x^\mu + 2 cy^\alpha \right) + \pdot{\gamma}v_{\alpha\beta}\left( b\tilde\delta^\alpha_\mu \dot x^\mu + cy^\alpha y^\beta\right) + a\pdot{\gamma}g_{\mu\nu}\dot x^\mu \dot x^\nu = 0
	\end{align}
	Contraction with $y^\gamma$ gives
	\begin{equation}\label{y=xdot1}
	y_\alpha\left( b\tilde\delta^\alpha_\mu \dot x^\mu  + 2 cy^\alpha\right) = 0
	\end{equation}
	where we assumed the homogeneity conditions $g_{\mu\nu}(x,ky) = g_{\mu\nu}(x,y), v_{\alpha\beta}(x,ky) = v_{\alpha\beta}(x,y), k>0$. A general solution for \eqref{y=xdot1} is
	\begin{equation}
	b\tilde\delta^\alpha_\mu \dot x^\mu = -2cy^\alpha \Leftrightarrow y^\alpha = -\frac{b}{2c}\tilde\delta^\alpha_\mu \dot x^\mu 
	\end{equation}
	We can redefine the parameter on the curve as $s\rightarrow s' = -\frac{2c}{b}s$ so that the last relation reads
	\begin{equation}\label{y=xdot}
	y^\alpha = \tilde\delta^\alpha_\mu \dot x^\mu 
	\end{equation}
	where we have redifined $\dot x = dx/ds'$.
	
	The Lagrangian \eqref{full lagrangian} with the new definitions can be written in the following equivalent form:
	\begin{equation}\label{new lagrangian}
	L(x,\dot x,y) = \left[-g_{\mu\nu}\dot x^\mu \dot x^\nu + z \left(- 2 \tilde \delta^\alpha_\mu v_{\alpha\beta}\dot x^\mu y^\beta + v_{\alpha\beta}y^\alpha y^\beta \right) \right]^{1/2}
	\end{equation}
	with $z = -b^2/4ac$. The remaining Euler-Lagrange equations \eqref{E-L x} along with \eqref{y=xdot} give the curve equation
	\begin{equation}\label{genpath}
	\left(\delta^\nu_\kappa + z g^{\nu\lambda}\tilde\delta^\alpha_\kappa \delta^\beta_\lambda v_{\alpha\beta}\right)\ddot x^\kappa + \left(\gamma^\nu_{\kappa\lambda} + z\sigma^\nu_{\kappa\lambda} \right) \dot x^\kappa \dot x^\lambda = 0
	\end{equation}
	with 
	\begin{equation}
	\sigma^\nu_{\kappa\lambda} = \frac{1}{2} g^{\mu\nu}\left(\tilde\delta^\alpha_\mu \tilde\delta^\beta_\lambda \partial_\kappa v_{\alpha\beta} + \tilde\delta^\alpha_\mu \tilde\delta^\beta_\kappa \partial_\lambda v_{\alpha\beta} - \tilde\delta^\alpha_\kappa \tilde\delta^\beta_\lambda \partial_\mu v_{\alpha\beta} \right)
	\end{equation}
	where we have set the normalization condition
	\begin{equation}\label{normcond}
	\frac{d}{ds} L = 0 \Leftrightarrow \frac{d}{ds} \left[ \left(g_{\mu\nu} + z \tilde\delta^\alpha_\mu \delta^\beta_\nu v_{\alpha\beta}\right) \dot x^\mu \dot x^\nu\right] = 0
	\end{equation}
	From \eqref{normcond} we get
	\begin{equation}
	\int_{x_0}^{x}\sqrt{\overline g_{\mu\nu} dx^\mu dx^\nu} = \lambda(s-s_0)
	\end{equation}
	with $\lambda$ integration constant and we have set $\overline g_{\mu\nu} = g_{\mu\nu} + z \tilde\delta^\alpha_\mu \delta^\beta_\nu v_{\alpha\beta} $.	
	This condition constrains the parameter along the curves up to affine transformations.
	
	{\bf Remark:} Relations \eqref{y=xdot} and \eqref{genpath} do not describe curves of stationary lenght with respect to Sasaki-type metric \eqref{bundle metric}. However, in the limit $z \rightarrow 0$, relation \eqref{genpath} converges to the geodesics equation \eqref{geodesics}, which is a stationary curve with respect to the horizontal metric $g_{\mu\nu}$.
	
	\subsection{Curves from two distinct Lagrangians}
	
	In this paragraph, we study the two distinct sets of curves that come from two Lagrangians $ L_H=\sqrt{-g_{\mu\nu}\dot x^\mu x^\nu} $ and $ L_V=\sqrt{\tilde \delta^\alpha_\mu v_{\alpha\beta}\dot x^\mu y^\beta + c v_{\alpha\beta}y^\alpha y^\beta} $. The first is derived from \eqref{full lagrangian} by keeping just the horizontal term and the second is derived by keeping the vertical and interaction terms.
	
	Euler-Lagrange equations for $L_H$ give:
	\begin{gather}
	\ddot x^\nu + \gamma^\nu_{\kappa\lambda}\dot x^\kappa \dot x^\lambda + 2 C^\nu_{\lambda\beta}\dot x^\lambda \dot y^\beta = 0 \label{LH1} \\
	\pdot{\alpha} g_{\mu\nu} \dot x^\mu \dot x^\nu = 0 \label{LH2}
	\end{gather}
	and respectively for $L_V$ they give:
	\begin{gather}
	\partial_\mu v_{\alpha\beta}\tilde \delta^\alpha_\kappa \dot x^\kappa y^\beta - \(\partial_\kappa v_{\alpha\beta}\dot x^\kappa + \pdot{\gamma}v_{\alpha\beta}\dot y^\gamma \)\tilde\delta^\alpha_\mu y^\beta  - v_{\alpha\beta}\tilde\delta^\alpha_\mu\dot y^\beta  = 0 \label{LH3}\\
	y^\beta = -\frac{1}{2c}\tilde \delta^\beta_\mu\dot x^\mu \label{LH4}
	\end{gather}
	We redefine the parameter along the curves as $s\rightarrow s' = -2c s$ so that \eqref{LH4} reads
	\begin{equation}
	y^\beta = \tilde \delta^\beta_\mu \dot x^\mu
	\end{equation}
	In the above cases, we have set $L_H = L_V = 1$. We note that the first set of equations \eqref{LH1}, \eqref{LH2} can accept the partial solution $y^\alpha = \tilde\delta^\alpha_\mu \dot x^\mu$, which does not come directly from the Lagrangian $L_H$. Using this solution, \eqref{LH1} becomes
	\begin{equation}
	\ddot x^\nu + \gamma^\nu_{\kappa\lambda}\dot x^\kappa \dot x^\lambda = 0
	\end{equation}
	due to homogeneity of $C^\nu_{\lambda\alpha}$ on $y$, while \eqref{LH2} is identically verified.
	
	\subsection{Geodesics}
	In this section we will derive the geodesics equations on $TM$, i.e. the curves of stationary length with respect to the metric \eqref{bundle metric}:	
	\begin{equation}	
	\mathcal{G} = g_{\mu\nu}(x,y)\, \de x^{\mu}\otimes \de x^{\nu} + v_{\alpha\beta}(x,y)\, \delta y^{\alpha}\otimes \delta y^{\beta}
	\end{equation}
	\vspace{5mm}
	The Lagrangian is written as
	\begin{equation}\label{geodesics lagrangian}
	\mathcal{L} = \sqrt{g_{\mu\nu}\dot x^\mu \dot x^\nu + v_{\alpha\beta}\frac{\delta y{^\alpha}}{\delta s}\frac{\delta y{^\beta}}{\delta s}} 
	\end{equation}
	with $\dot x^\mu = dx^\mu/ds$, where s is the arclength and
	\begin{equation}
	\frac{\delta y^\alpha}{\delta s} = \dot y^a  + N^{\alpha}_k\dot x^k 
	\end{equation}
	So the Lagrangian can be written in the form:
	\begin{equation}\label{geodesics full lagrangian}
	\mathcal{L}=\sqrt{\(g_{\mu\nu} + v_{\alpha\beta}N^{\alpha}_{\mu}N^{\beta}_{\nu}\)\dot x^\mu \dot x^\nu + 2v_{\alpha\beta}\dot y^\alpha N^{\beta}_{\lambda}\dot x^\lambda +v_{\alpha\beta}\dot y^\alpha \dot y^\beta }
	\end{equation}
	The Euler-Lagrange equations are given by:
	\begin{equation}\label{ELx}
	\pder{\mathcal{L}}{x^\rho} = \frac{d}{ds}
	\frac{\partial \mathcal{L}}{\partial \dot x^\rho } 
	\end{equation}
	and
	\begin{equation}\label{ELy}
	\pder{\mathcal{L}}{y^\gamma}=\frac{d}{ds}\pder{\mathcal{L}}{\dot y^\gamma }
	\end{equation}
	We define :   
	\begin{equation}
	h_{\mu\nu} = v_{\alpha\beta}N^{\alpha}_{\mu}N^{\beta}_{\nu}
	\end{equation}
	And we calculate the terms required below: 
	\begin{align}
	\frac{\partial \mathcal L}{\partial x^\rho} = &\, \frac{1}{2 \mathcal L}\big[\(\partial_{\rho}g_{\mu\nu} + \partial_{\rho}h_{\mu\nu}\)\dot x^\mu \dot x^\nu  + 2\partial_{\rho}v_{\alpha\beta}N^{\beta}_{\lambda}\dot y^\alpha \dot x^{\lambda}  +  2v_{\alpha\beta}\partial_{\rho}N^{\beta}_{\lambda}\dot y^\alpha \dot x^\lambda  + \partial_{\rho}v_{\alpha\beta}\dot y^\alpha \dot y^\beta \big] \label{pLx}\\
	\frac{\partial \mathcal L}{\partial\dot  x^\rho } = &\, \frac{1}{2\mathcal L}\big[2\(g_{\mu\rho} + h_{\mu\rho}\)\dot x^\mu  + 2v_{\alpha\beta}N^\beta_\rho\dot y^\alpha \big] \label{pLxdot}\\
	\frac{\partial \mathcal L}{\partial y^\gamma} = &\, \frac{1}{2\mathcal L}\Big[\(\dot \partial_{\gamma} g_{\mu\nu} + \dot \partial_{\gamma} h_{\mu\nu}\)\dot x^\mu \dot x^\nu  + 2 \dot \partial_{\gamma} \(v_{\alpha\beta}N^\beta_\lambda\)\dot x^\lambda  \dot y^\alpha 
    + \dot \partial_{\gamma} v_{\alpha\beta}\dot y^\alpha \dot y^\beta \Big] \label{pLy} \\
	\frac{d}{ds}\frac{\partial \mathcal L}{\partial\dot  y^\gamma } = &\, \partial_\kappa \(v_{\beta\gamma}N^\beta_\lambda\)\dot{x^\kappa}\dot x^\lambda  + \dot \partial_\alpha \(v_{\beta\gamma}N^\beta_\lambda\)\dot y^\alpha \dot x^\lambda + \partial_\kappa v_{\beta\gamma}\dot y^\beta \dot x^\kappa  + \dot{\partial_\alpha}v_{\beta\gamma}\dot y^\beta \dot y^\alpha  + v_{\beta\gamma}\ddot y^\beta \label{pLydot}
	\end{align}
	Where we normalized $\mathcal{L}$ to 1.
	
	Taking the relations \eqref{pLy}, \eqref{pLydot} and substituting them to \eqref{ELy} we get the geodesics equation of the vertical space:
	\begin{align}
	\ddot y^\delta + C^{\delta}_{\alpha\beta} \dot y^\alpha \dot y^\beta = \frac{1}{2}v^{\gamma\delta}\(\pdot{\gamma}g_{\mu\nu}+\pdot{\gamma}h_{\mu\nu}\) \dot x^\mu \dot x^\nu + v^{\gamma\delta}\pdot{\gamma} \(v_{\alpha\beta}N^{\beta}_{\lambda}\)\dot y^\alpha \dot x^\lambda - v^{\gamma\delta}\partial_{\kappa}v_{\beta\gamma} \dot x^\kappa \dot y^\beta \label{ygeodesics}
	\end{align}
	Now, if we consider $\dot x = 0$ in our space, we get a subspace of $TM$ which is called the tangent Riemannian space $T_xM$. It has been shown that in $T_xM$ the geodesics equation has the form \cite{stavrinos1992tidal}:
	\begin{equation}\label{geodesics tangent}
	\ddot y^\delta + C^{\delta}_{\alpha\beta} \dot y^\alpha \dot y^\beta = 0
	\end{equation}
	We observe that we get the same equation if we set $\dot x = 0$ in \eqref{ygeodesics}. This means that in our metric space the Lagrangian \eqref{geodesics full lagrangian} provides us with a set of generalized geodesics equations. 
	
	Like before, by taking \eqref{pLx}, \eqref{pLxdot} and substituting them to \eqref{ELx} we get the geodesics equation for the horizontal space.
	\begin{align}
	\ddot x^\kappa + \gamma^{\kappa}_{\mu\lambda}\dot{x^\mu}\dot x^\lambda = &\, \frac{1}{2}g^{\rho\kappa}\partial_{\rho}h_{\mu\nu}\dot x^\mu \dot x^\nu
	+g^{\rho\kappa}\partial_{\rho}\(v_{\alpha\beta}N^{\beta}_{\lambda}\)\dot x^\lambda \dot y^\alpha + \frac{1}{2}g^{\rho\kappa}\partial_{\rho}v_{\alpha\beta}\dot y^\alpha \dot y^\beta - g^{\rho\kappa}\partial_{\lambda}h_{\mu\rho}\dot x^\mu \dot x^\lambda - g^{\rho\kappa}h_{\mu\rho}\ddot x^\mu \nonumber\\ 
	&-g^{\rho\kappa}\(\pdot{\gamma} g_{\mu\rho}+\pdot{\gamma}
	h_{\mu\rho}\)\dot x^\mu \dot{y^\gamma} - g^{\rho\kappa}\ddot y^\alpha v_{\alpha\beta}N^{\beta}_{\rho} -g^{\rho\kappa}\partial_{\lambda}\(v_{\alpha\beta}N^{\beta}_{\rho}\)\dot x^\lambda\dot y^\alpha -g^{\rho\kappa}\pdot{\gamma}\(v_{\alpha\beta}N^{\beta}_{\rho}\)\dot y^\gamma \dot y^\alpha \label{xgeodesics}
	\end{align}
	where we have used $\gamma^{\kappa}_{\mu\lambda}$ which are the Christoffel symbols for the metric $g_{\mu\nu}(x,y)$.\\  
	{\bf Remark:} We can find the necessary conditions so that the Lagrangian in \eqref{geodesics lagrangian} and the Lagrangian in \eqref{full lagrangian} are equal. These conditions, after some calculations, are shown below:
	\begin{equation}\label{cond1}
	2 \dot y^\alpha N^{\gamma}_{\nu}=b\tilde\delta^{\alpha}_{\nu}y^{\gamma}
	\end{equation}
	\begin{equation}\label{cond2}
	b^{2}y^{\gamma}y^{\delta}b\, \tilde \delta^{\alpha}_{\nu} b\tilde \delta^{\beta}_{\mu} = 4c\, y^{\alpha}y^{\beta}N^{\gamma}_{\nu}N^{\delta}_{\mu}
	\end{equation}
	This system has the trivial solution $y^\alpha = \dot y^\alpha = 0$. For $N^\gamma_\nu \neq 0, y^\alpha\neq 0, \dot y^\alpha \neq 0, b\neq 0$ and $c>0$, if we substitute \eqref{cond1} to \eqref{cond2} we get
	\begin{equation}\label{cond3}
	\frac{b}{\sqrt{c}}\dot y^\alpha \tilde \delta^\beta_\mu = y^\beta \tilde \delta^\alpha_\mu
	\end{equation}
	For \eqref{cond3} to have nontrivial solutions, it must be that $\tilde \delta^\alpha_\mu = \tilde \delta^\beta_\mu = 1 \Leftrightarrow \alpha = \beta = \mu+4$. Under this condition, \eqref{cond3} gives
	\begin{equation}\label{cond4}
	\dot y^\alpha = \lambda y^\alpha
	\end{equation}
	where $\lambda = b/\sqrt{c}$. We can solve \eqref{cond4} to get
	\begin{equation}\label{condf}
	y^\alpha = y^\alpha_0 e^{\lambda s + k}
	\end{equation}
	with $y^\alpha_0$ and $k$ integration constants.
	

	\section{Field equations for an Einstein-Hilbert-like action}\label{sec: field equations}

	In a previous work, field equations on the Lorentz tangent bundle have been derived from a proper action \cite{Triantafyllopoulos:2018bli}. In this paragraph, we use an extension of Stokes theorem on the Lorentz tangent bundle to derive field equations for the metric and the nonlinear connection.
	
	An Einstein-Hilbert-like action with a matter sector on the Lorentz tangent bundle is defined as
	\begin{equation}
	K = \int_{\mathcal N} d^8\mathcal U \sqrt{|\Gd|}\, \R + 2\kappa \int_{\mathcal N} d^8\mathcal U \sqrt{|\Gd|}\,\mathcal L_M
	\end{equation}
	for some closed subspace $\mathcal N\subset TM$, where $|\Gd|$ is the absolute value of the metric determinant, $\mathcal L_M$ is the Lagrangian of the matter fields, $\kappa$ is a constant and
	\begin{equation}
	d^8\mathcal U = \de x^0 \wedge \ldots \wedge\de x^3 \wedge \de y^4 \wedge \ldots \wedge \de y^7
	\end{equation}
	Variation with respect to $g_{\mu\nu}$, $v_{\alpha\beta}$ and $N^\alpha_\kappa$ gives
	\begin{align}
	\Delta K = &\, \int_{\mathcal N} d^8 \mathcal U (\overline R + S) \Delta \sqrt{|\Gd|} + \int_{\mathcal N} d^8 \mathcal U \sqrt{|\Gd|} (\Delta \overline R + \Delta S) 
 + \, 2\kappa\int_{\mathcal N} d^8\mathcal U \,\Delta\!\(\sqrt{|\Gd|}\,\mathcal L_M\) \label{DK}
	\end{align}
	with
	\begin{align}
	\Delta\sqrt{|\Gd|} = & -\frac{1}{2}\sqrt{|\Gd|}\left( g_{\mu\nu}\Delta g^{\mu\nu} + v_{\alpha\beta}\Delta v^{\alpha\beta}\right) \label{DG}\lin
	\Delta \overline{R} = & \, 2g^{\mu[\kappa}\pdot{\alpha}L^{\nu]}_{\mu\nu} \Delta N^\alpha_\kappa + \overline{R}_{\mu\nu}\Delta g^{\mu\nu} + \mathcal D_\kappa Z^\kappa \label{DR} \lin
	\Delta S = &\, S_{\alpha\beta} \Delta v^{\alpha\beta} + \mathcal D_\gamma B^\gamma \label{DS}
	\end{align}
	where $\overline R_{\mu\nu} = R_{(\mu\nu)} + \Omega^\alpha_{\kappa(\mu} C^\kappa_{\nu)\alpha}$, $\overline R \equiv g^{\mu\nu}\overline R_{\mu\nu} = R$ and
	\begin{align}
	Z^\kappa = &\, g^{\mu\nu}\Delta L^\kappa_{\mu\nu} - g^{\mu\kappa} \Delta L^\nu_{\mu\nu} = -\D_\nu\Delta g^{\nu\kappa} + g^{\kappa\lambda}g_{\mu\nu}\D_\lambda \Delta g^{\mu\nu} + 2\( g^{\kappa\mu}C^\lambda_{\lambda\alpha} - g^{\kappa\lambda}C^\mu_{\lambda\alpha} \) \Delta N^\alpha_\mu \label{Zdef}\\
	B^\gamma = &\, v^{\alpha\beta}\Delta C^\gamma_{\alpha\beta} - v^{\alpha\gamma}\Delta C^\beta_{\alpha\beta} 
	=  -\D_\alpha \Delta v^{\alpha\gamma} + v^{\gamma\delta}v_{\alpha\beta} \D_\delta\Delta v^{\alpha\beta} \label{Bdef}
	\end{align}
	
	The extension of Stokes theorem on the Lorentz tangent bundle is given by the relations
	\begin{align}
	\int_{\mathcal N} d^8 \mathcal{U}\sqrt{|\Gd|}\,\mathcal D_\mu H^\mu = & \int_{\mathcal N} d^8 \mathcal{U}\sqrt{|\Gd|}\,\mathcal T^\alpha_{\mu\alpha}H^\mu + \int_{\partial {\mathcal N}} n_\mu H^\mu \mathcal{\tilde E} \label{stokesh}\\
	\int_{\mathcal N} d^8 \mathcal{U}\sqrt{|\Gd|}\,\mathcal D_\alpha W^\alpha = &  -\int_{\mathcal N} d^8 \mathcal{U}\sqrt{|\Gd|}\, C^\mu_{\mu\alpha}W^\alpha + \int_{\partial {\mathcal N}}n_\alpha W^\alpha \mathcal{\tilde E} \label{stokesv}
	\end{align}
	where $H = H^\mu \delta_\mu$ and $W = W^\alpha \pdot{\alpha}$ are vector fields on $TM$, $\tilde{\mathcal E}$ is the Levi-Civita tensor on the boundary $\partial {\mathcal N}$, $(n_\mu,n_\alpha)$ is the normal vector on the boundary and $\mathcal T^\alpha_{\mu\beta} = \pdot{\beta}N^\alpha_\mu - L^\alpha_{\beta\mu}$. For more details see Appendix \ref{sec: stokes}. Using relation \eqref{stokesh} and eliminating boundary terms we get
	\begin{align}
	\int_{\mathcal N} d^8 \mathcal{U}\sqrt{|\Gd|}\,\mathcal D_\kappa Z^\kappa = &\,  \int_{\mathcal N} d^8 \mathcal{U}\sqrt{|\Gd|}\,\mathcal T^\alpha_{\kappa\alpha}Z^\kappa \nonumber\\
	= &\, \int_{\mathcal N} d^8 \mathcal{U}\sqrt{|\Gd|}\,\D_\nu \left[ \mathcal T^\beta_{\kappa\beta} \( -\Delta g^{\nu\kappa} + g^{\nu\kappa}g_{\mu\lambda}\Delta g^{\mu\lambda}\)\right] 
	-\int_{\mathcal N} d^8 \mathcal{U}\sqrt{|\Gd|}\,\left[ -\D_\nu\mathcal T^\beta_{\mu\beta} + g_{\mu\nu}\D^\lambda\mathcal T^\beta_{\lambda\beta}\right]\Delta g^{\mu\nu} \nonumber\\
	& + 2\int_{\mathcal N} d^8 \mathcal{U}\sqrt{|\Gd|}\,\mathcal T^\beta_{\kappa\beta}\( g^{\kappa\mu}C^\lambda_{\lambda\alpha} - g^{\kappa\lambda}C^\mu_{\lambda\alpha}\)\Delta N^\alpha_\mu
	\end{align}
	where we have used the Leibniz rule for the covariant derivative. Using \eqref{stokesh} again and eliminating the new boundary terms we get
	\begin{align}
	\int_{\mathcal N} d^8 \mathcal{U}\sqrt{|\Gd|}\,\mathcal D_\kappa Z^\kappa 
	= \int_{\mathcal N} d^8 \mathcal{U}\sqrt{|\Gd|}\,\left(\delta^{(\lambda}_\nu\delta^{\kappa)}_\mu - g^{\kappa\lambda}g_{\mu\nu} \right)\left(\mathcal D_\kappa\mathcal T^\beta_{\lambda\beta} - \mathcal T^\gamma_{\kappa\gamma}\mathcal T^\beta_{\lambda\beta}\right) \Delta g^{\mu\nu} + \int_{\mathcal N} d^8 \mathcal{U}\sqrt{|\Gd|}\,4\mathcal{T}^\beta_{\kappa\beta}g^{\kappa[\mu}C^{\lambda]}_{\lambda\alpha} \Delta N^\alpha_\mu \label{DZ}
	\end{align}
	Similarly, using relation \eqref{stokesv} and eliminating the boundary terms we get
	\begin{align}
	\int_{\mathcal N} d^8\mathcal U \sqrt{|\Gd|}\, \D_\alpha B^\alpha = &\, -\int_{\mathcal N} d^8\mathcal U \sqrt{|\Gd|}\,C^\mu_{\mu\beta}B^\beta \nonumber\\
	= &\, -\int_{\mathcal N} d^8\mathcal U \sqrt{|\Gd|}\,\D_\alpha \left[ C^\mu_{\mu\beta}\Delta v^{\alpha\beta} - v^{\alpha\beta}v_{\gamma\delta}C^\mu_{\mu\beta} \Delta v^{\gamma\delta}\right] \nonumber\\ 
	& - \int_{\mathcal N} d^8\mathcal U \sqrt{|\Gd|}\,\( \D_\alpha C^\mu_{\mu\beta} - v^{\gamma\delta}v_{\alpha\beta}\D_\gamma C^\mu_{\mu\delta}\)\Delta v^{\alpha\beta}
	\end{align}
	where again we used the Leibniz rule. Applying \eqref{stokesv} again and eliminating the new boundary terms we get
	\begin{equation}
	\int_{\mathcal N} d^8\mathcal U \sqrt{|\Gd|}\,\D_\alpha B^\alpha  = \int_{\mathcal N} d^8\mathcal U \sqrt{|\Gd|}\,\left(v^{\gamma\delta}v_{\alpha\beta} - \delta^{(\gamma}_\alpha\delta^{\delta)}_\beta \right)\left(\mathcal D_\gamma C^\mu_{\mu\delta} - C^\nu_{\nu\gamma}C^\mu_{\mu\delta} \right) \label{DB}
	\end{equation}
	Finally, combining equations (\ref{DK}-\ref{Bdef}), \eqref{DZ}, \eqref{DB} and setting $\Delta K = 0$ we get the following equations
	\begin{gather}
	{\overline R_{\mu\nu}} - \frac{1}{2}({\overline R}+{S})\,{g_{\mu\nu}} + \left(\delta^{(\lambda}_\nu\delta^{\kappa)}_\mu - g^{\kappa\lambda}g_{\mu\nu} \right)\left(\mathcal D_\kappa\mathcal T^\beta_{\lambda\beta} - \mathcal T^\gamma_{\kappa\gamma}\mathcal T^\beta_{\lambda\beta}\right) = \kappa {T_{\mu\nu}} \label{feq1}\lin
	{S_{\alpha\beta}} - \frac{1}{2}({\overline R}+{S})\,{v_{\alpha\beta}}  + \left(v^{\gamma\delta}v_{\alpha\beta} - \delta^{(\gamma}_\alpha\delta^{\delta)}_\beta \right)\left(\mathcal D_\gamma C^\mu_{\mu\delta} - C^\nu_{\nu\gamma}C^\mu_{\mu\delta} \right) = \kappa {Y_{\alpha\beta}} \label{feq2}\lin
	g^{\mu[\kappa}\pdot{\alpha}L^{\nu]}_{\mu\nu} +  2 \mathcal T^\beta_{\mu\beta}g^{\mu[\kappa}C^{\lambda]}_{\lambda\alpha} = \frac{\kappa}{2}\mathcal Z^\kappa_\alpha \label{feq3}
	\end{gather}
	with
	\begin{align}
	T_{\mu\nu} \equiv -\frac{2}{\sqrt{|\Gd|}}\frac{\Delta\left(\sqrt{|\Gd|}\,\mathcal{L}_M\right)}{\Delta g^{\mu\nu}}\label{em1}\\
	Y_{\alpha\beta} \equiv -\frac{2}{\sqrt{|\Gd|}}\frac{\Delta\left(\sqrt{|\Gd|}\,\mathcal{L}_M\right)}{\Delta v^{\alpha\beta}} \label{em2}\\
	\mathcal Z^\kappa_\alpha \equiv -\frac{2}{\sqrt{|\Gd|}}\frac{\Delta\left(\sqrt{|\Gd|}\,\mathcal{L}_M\right)}{\Delta N^\alpha_\kappa}\label{em3}
	\end{align}
	where $\delta^\mu_\nu$ and $ \delta^\alpha_\beta$ are the Kronecker symbols.
	
	We will make some comments in order to give some physical interpretation in relation to the equations \eqref{em1}, \eqref{em2} and \eqref{em3}. Lorentz violations produce anisotropies in the space and the matter sector \cite{kostelecky 2004,Kouretsis:2008ha}. These act as a source of local anisotropy and can contribute to the energy-momentum tensors of the horizontal and vertical space $T_{\mu\nu}$ and $Y_{\alpha\beta}$. As a result, the energy-momentum tensor $T_{\mu\nu}$ contains the additional information of local anisotropy of matter fields. $Y_{\alpha\beta}$, on the other hand, is an object with no equivalent in Riemannian gravity. It contains more information about local anisotropy which is produced from the metric $v_{\alpha\beta}$ which includes additional internal structure of space-time. Finally, the energy-momentum tensor $ \mathcal Z^\kappa_\alpha $ reflects the dependence of matter fields on the nonlinear connection $N^\alpha_\mu$, a structure which induces an interaction between internal and external spaces. This is different from $T_{\mu\nu}$ and $Y_{\alpha\beta}$ which depend on just the external or internal structure respectively.

	\section{Concluding Remarks}\label{sec: conclusion}
	In this work we study Lagrangians $ L(x,\dot x, y) $ that include directional variables $y$ and tangent vectors $\dot x$ on generalized paths $\(x(s), y(s)\)$, with $s$ an affine parameter, on a Lorentz tangent bundle $TM$ of a space-time manifold $M$. This Lagrangian includes interaction terms and a condition between $\dot x$ and $y$ is provided by this framework. In addition, the geodesic equations \eqref{xgeodesics}, \eqref{ygeodesics} on $TM$ with a normalized Lagrangian are compared with the generalized paths \eqref{y=xdot}, \eqref{genpath} of $L(x,\dot x,y)$ and we find the necessary conditions for $L(x,\dot x,y)$ and $\mathcal L(x,\dot x,y,\dot y)$ to be equal.
	
	Moreover, we derived a generalized form of Einstein-Hilbert field equations in the framework of a Lorentz tangent bundle of a 4-dim. space-time (\ref{feq1}-\ref{em3}) by using an extension of the Stokes theorem on the bundle. The third set of equations, namely \eqref{feq3}, \eqref{em3}, include the concept of nonlinear connection and its direct influence on the matter fields.
	
	A study of the gravitational field depending on the external (base space-time) and the internal structure (observers) can provide a wider framework of research of the gravitational field. Such a consideration could give predictions with extra terms/corrections beyond the classical observations of general relativity and of standard cosmology. This can constitute an interesting study for our next work.
	
 \appendix

 \section{Stokes theorem on the Lorentz tangent bundle}\label{sec: stokes}
	
	To eliminate terms of the form $\int_{\mathcal N} d^8\mathcal U \sqrt{|\Gd|}\, \mathcal D_A V^A$, we need an extension of Stokes theorem on the Lorentz tangent bundle. In its general form, the theorem reads:
	\begin{equation}\label{stokes}
	\int_\mathcal N\de\omega = \int_{\partial \mathcal N}\omega
	\end{equation}
	where $\mathcal N$ is a closed subspace and $\omega$ a $(n-1)$-form of an $n-$dimensional manifold. In our case, the role of the manifold is played by the $ 8- $dimensional tangent bundle, so $n=8$.
	
	We begin by calculating $\de\omega$ on the adapted basis. The properties  of the exterior derivative are
	\begin{enumerate}
		\item $ \de f(Y) = Y(f) $, with $ f(\mathcal U)$ a real function and $Y$ a vector field on $TM$.
		\item $ \de (\omega \wedge \psi) = \de \omega \wedge \psi + (-1)^p \omega \wedge \de\psi $, with $ \omega: p-$form and $\psi: q-$form.
		\item $ \de(\omega + \psi) = \de\omega + \de\psi $
		\item $ \de^2\omega = 0 $
	\end{enumerate}
	For the $p-$form
	\begin{align}
	\omega  = \omega_{A_1 A_2 \ldots A_p}E^{A_1}\otimes E^{A_2}\otimes\ldots\otimes E^{A_p} = \frac{1}{p!}\omega_{A_1 A_2 \ldots A_p} E^{A_1}\wedge E^{A_2}\wedge\ldots\wedge E^{A_p}
	\end{align}
	we get
	\begin{align}
	\de\omega = & \,\frac{1}{p!} \left[ E_A\omega_{A_1 A_2 \ldots A_p}E^A\wedge E^{A_1}\wedge E^{A_2}\wedge\ldots\wedge E^{A_p} \right. + \omega_{A_1 A_2 \ldots A_p} \de E^{A_1}\wedge E^{A_2}\wedge\ldots\wedge E^{A_p} \nonumber\\
	& - \omega_{A_1 A_2 \ldots A_p}E^{A_1}\wedge \de E^{A_2}\wedge\ldots\wedge E^{A_p} + \ldots  \left. + (-1)^{p-1}\omega_{A_1 A_2 \ldots A_p}E^{A_1}\wedge E^{A_2}\wedge\ldots\wedge \de E^{A_p}\right] \label{dw1}
	\end{align}
	where we used the properties of the exterior derivative. We find
	\begin{align}\label{dE}
	\de E^A & = \delta^A_\mu\de^2 x^\mu + \delta^A_\alpha \de\delta y^\alpha \nonumber\\
	& = \delta^A_\alpha \left( \de^2 y^\alpha + \de N^\alpha_\mu \wedge \de x^\mu + N^\alpha_\mu \de^2x^\mu \right) \nonumber\\
	& = \delta^A_\alpha \left( \delta_\nu N^\alpha_\mu \de x^\nu \wedge \de x^\mu + \pdot{\beta} N^\alpha_\mu \delta y^\beta \wedge \de x^\mu \right)
	\end{align}
	Additionally, we write
	\begin{align}\label{Ew}
	E_B\,\omega_{A_1 \ldots A_p}  = \mathcal D_B\, \omega_{A_1 \ldots A_p} + \Gamma^A_{A_1 B}\,\omega_{A \ldots A_p} + \ldots + \Gamma^A_{A_p B}\, \omega_{A_1 \ldots A}
	\end{align}
	From \eqref{dw1}, \eqref{dE} and \eqref{Ew} we get
	\begin{align}
	p!\,\de\omega = &\, \left( \mathcal D_{[B}\,\omega_{A_1 \ldots A_p]} + \Gamma^A_{[A_1B}\,\omega_{A \ldots A_p]} + \ldots +\Gamma^A_{[A_p B}\,\omega_{A_1 \ldots A]}\right) \wedge E^B\wedge E^{A_1}\wedge \ldots \wedge E^{A_p} \nonumber\\
	& + p\,\omega_{\alpha A_1 \ldots A_{p-1}} \left( \delta_\nu N^\alpha_\mu\de x^\nu \wedge \de x^\mu + \pdot{\beta}N^\alpha_\mu \delta y^\beta \wedge \de x^\mu\right) \wedge E^{A_1} \wedge\ldots \wedge E^{A_p-1} \label{dw2}
	\end{align}
	Now, the Hodge dual of a $q$-form $\psi$ is defined as the $(n-q)$-form:
	\begin{equation}\label{hodgedual}
	\(\star\,\psi\)_{M_1\ldots M_{n-q}} = \frac{1}{q!} \tensor{\mathcal E}{^{N_1\ldots N_q}_{M_1\ldots M_{n-q}}}\tensor{\psi}{_{N_1\ldots N_q}}
	\end{equation}
	where $\star$ is the Hodge duality operator and $\mathcal E$ the Levi-Civita tensor:
	\begin{align}\label{levicivita}
	\mathcal E & = \sqrt{|\Gd|} E^0\wedge\ldots \wedge E^7 \nonumber \\
	& = \sqrt{|\Gd|}\, \de x^0 \wedge \ldots \wedge\de x^3 \wedge \de y^4 \wedge \ldots \wedge \de y^7 \nonumber\\
	& =  \sqrt{|\Gd|}\, d^8\mathcal U
	\end{align}
	In \eqref{stokes}, $\omega$ is a $n-1$-form, so it is the Hodge dual of a one form $V = V_NE^N$ of $TM$. From \eqref{hodgedual} we get 
	\begin{equation}\label{w=V}
	\omega_{A_1 \ldots A_p} = (\star V)_{A_1 \ldots A_p} = \tensor{\mathcal E}{^N_{A_1\ldots A_p}}V_N = \mathcal E_{NA_1\ldots A_p}V^N
	\end{equation}
	From relations \eqref{dw2} and \eqref{w=V} we get
	\begin{align}
	(\de\omega)_{BA_1\ldots A_p} = &\, (\de\star V)_{BA_1\ldots A_p} \nonumber\\
	= &\, (p+1)\left( \mathcal E_{N[A_1\ldots A_p}\mathcal D_{B]} V^N + \mathcal E_{NA[A_1\ldots A_p}\Gamma^A_{A_1B]}V^N\right. \nonumber\\
	& + \ldots \nonumber \\
	& + \left.\mathcal E_{N[A_1\ldots|A|}\Gamma^A_{A_pB]} V^N + p\,\mathcal  E_{NA[A_2\ldots A_p} K^A_{A_1B]}V^N \right) \label{dw3}
	\end{align}
	with 
	\begin{equation}\label{Kabm}
	K^\alpha_{\mu\nu} = \delta_\nu N^\alpha_\mu,\quad K^\alpha_{\mu\beta} = \pdot{\beta} N^\alpha_\mu
	\end{equation}
	and $K^A_{BM} = 0$ for every other combination of indices. From \eqref{w=V} and for $p=n-1$ we find the Hodge dual of expression \eqref{dw3} as the zero-form field
	\begin{align}
	\star\, \de\omega = &\, \frac{1}{n!}\mathcal{E}^{BA_1\ldots A_{n-1}}(\de\star V)_{BA_1\ldots A_{n-1}} \nonumber\\
	= & \,\frac{(-1)^s}{n!} n\big[ (n-1)!\, \D_NV^N  - (n-1)2!(n-2)!\, \delta^{[C}_N\delta^{B]}_MV^N\(\Gamma^M_{CB} + K^M_{CB}\)\big] \nonumber \\
	= &\, \D_N V^N - 2 \( \Gamma^M_{[NM]} + K^M_{[NM]}\)V^N \label{dw4}
	\end{align}
	where in the second step we used the identity
	\begin{equation}
	\mathcal{E}^{M_1\ldots Mq A_1\ldots A_{n-q}}\mathcal{E}_{M_1\ldots Mq B_1\ldots B_{n-q}}
	= (-1)^sq!(n-q)!\,\delta^{[A_1}_{B_1} \ldots \delta^{A_{n-q}]}_{B_{n-q}} \label{lciden}
	\end{equation}
	with $s$ the signature of the metric, in the case of \eqref{bundle metric} we have $s=2$. Now, $\star\, \de\omega$ is a zero form field or equivalently a scalar field, let's call it $\Phi$. Acting on it with $\star$ gives
	\begin{equation}\label{starf1}
	\star\Phi = \star\star \de\omega = \de\omega
	\end{equation}
	where in the last equality we used definition \eqref{hodgedual} and the identity \eqref{lciden}. From \eqref{hodgedual} and \eqref{levicivita} we also get
	\begin{equation}\label{starf2}
	\star\Phi = \Phi \mathcal E_{M_1\ldots M_N} E^{M_1}\otimes \ldots \otimes E^{M_n} = \Phi\mathcal E = \Phi\sqrt{|\Gd|}\, d^8\mathcal U
	\end{equation}
	Acting on \eqref{dw4} with $\star$ and using \eqref{starf1} and \eqref{starf2} we get
	\begin{equation}\label{dw}
	\de \omega = \left[ \D_N V^N - 2 \( \Gamma^M_{[NM]} + K^M_{[NM]}\)V^N \right] \sqrt{|\Gd|}\, d^8\mathcal U
	\end{equation}
	On the other hand, it is a known fact that
	\begin{equation}\label{intw}
	\int_{\partial\mathcal N} \omega = \int_{\partial\mathcal N} n_M V^M \mathcal{\tilde E} 
	\end{equation}
	with $\mathcal{\tilde E} $ the Levi-Civita tensor on the boundary and $n_M$ a normal vector field on the boundary. Putting together equations \eqref{stokes}, \eqref{dw} and \eqref{intw} we get
	\begin{equation}\label{stokestm}
	\int_{\mathcal N} d^8 \mathcal{U}\sqrt{|\Gd|}\,\mathcal D_N V^N = \int_{\partial\mathcal N} n_M V^M \mathcal{\tilde E} + 2\int_{\mathcal N} d^8 \mathcal{U}\sqrt{|\Gd|}\,\( \Gamma^M_{[NM]} + K^M_{[NM]}\)V^N
	\end{equation}
	Finally, for a canonical and distinguished $d-$connection (\ref{metric d-connection 1}-\ref{metric d-connection 4}), relations \eqref{Kabm} and \eqref{stokestm} give
	\begin{align}
	\int_{\mathcal N} d^8 \mathcal{U}\sqrt{|\Gd|}\,\mathcal D_\mu H^\mu = & \int_{\mathcal N} d^8 \mathcal{U}\sqrt{|\Gd|}\,\(\pdot{\alpha}N^\alpha_\mu - L^\alpha_{\alpha\mu}\)H^\mu + \int_{\partial {\mathcal N}} n_\mu H^\mu \mathcal{\tilde E} \\
	\int_{\mathcal N} d^8 \mathcal{U}\sqrt{|\Gd|}\,\mathcal D_\alpha W^\alpha = & -\int_{\mathcal N} d^8 \mathcal{U}\sqrt{|\Gd|}\, C^\mu_{\mu\alpha}W^\alpha + \int_{\partial {\mathcal N}}n_\alpha W^\alpha \mathcal{\tilde E} 
	\end{align}
	for the vector fields $H = H^\mu \delta_\mu$ and $W = W^\alpha \pdot{\alpha}$.

	\section*{Acknowledgments}
	This research is co-financed by Greece and the European Union (European Social Fund-ESF) through the Operational Programme ``Human Resources Development, Education and Lifelong Learning'' in the context of the project ``Strengthening Human Resources Research Potential via Doctorate Research'' (MIS-5000432), implemented by the State Scholarships Foundation (IKY). We would like to thank the unknown referee for his valuable comments on our work. We also want to sincerely thank Mr. Christos Savopoulos for the valuable discussions on our work.

\end{document}